\documentclass[aps,pra,twocolumn,superscriptaddress]{revtex4-2}

\usepackage{amsthm}
\usepackage{amsmath,bm}
\usepackage{amssymb}
\usepackage{amsfonts}
\usepackage{graphicx}
\usepackage{txfonts}
\usepackage{xcolor}
\usepackage{float}
\usepackage{braket}

\usepackage[colorlinks=true,linkcolor=blue,citecolor=blue,urlcolor=blue]{hyperref}

\newcommand{\eg}{\textit{e.g. }}

\newcommand{\avg}[1]{\langle#1\rangle}		% average
\newcommand{\var}{\text{Var}}	
	
\newcommand{\cov}{\text{Cov}}

\newcommand{\ad}{{a^\dagger}}

\newcommand{\op}[1]{\hat{#1}}				% operator with hat
\newcommand{\Vmin}{V_\text{min}}
\newcommand{\topt}{t_\text{opt}}

\newcommand{\ie}{\textit{i.e.} }

\begin{document}

\title{Quantum metrology with a squeezed Kerr oscillator}

\author{Jiajie Guo}
\affiliation{State Key Laboratory for Mesoscopic Physics, School of Physics, Frontiers Science Center for Nano-optoelectronics, Peking University, Beijing 100871, China}

\author{Qiongyi He}
\email{qiongyihe@pku.edu.cn}
\affiliation{State Key Laboratory for Mesoscopic Physics, School of Physics, Frontiers Science Center for Nano-optoelectronics, Peking University, Beijing 100871, China}
\affiliation{Collaborative Innovation Center of Extreme Optics, Shanxi University, Taiyuan, Shanxi 030006, China}
\affiliation{Peking University Yangtze Delta Institute of Optoelectronics, Nantong 226010, Jiangsu, China}  
\affiliation{Hefei National Laboratory, Hefei 230088, China}

\author{Matteo Fadel}
\email{fadelm@phys.ethz.ch}
\affiliation{Department of Physics, ETH Z\"{urich}, 8093 Z\"{urich}, Switzerland}

\begin{abstract}
We study the squeezing dynamics in a Kerr-nonlinear oscillator, and quantify the metrological usefulness of the resulting states. Even if the nonlinearity limits the attainable squeezing by making the evolution non-Gaussian, the states obtained still have a high quantum Fisher information for sensing displacements. However, contrary to the Gaussian case, the amplitude of the displacement cannot be estimated by simple quadrature measurements. Therefore, we propose the use of a measurement-after-interaction protocol where a linear quadrature measurement is preceded by an additional nonlinear evolution, and show the significant sensitivity enhancement that can be obtained. Our results are robust when considering realistic imperfections such as energy relaxation, and can be implemented in state-of-the-art experimental setups.
\end{abstract}

\maketitle

\begin{figure}[t]
    \begin{center}
	\includegraphics[width=85mm]{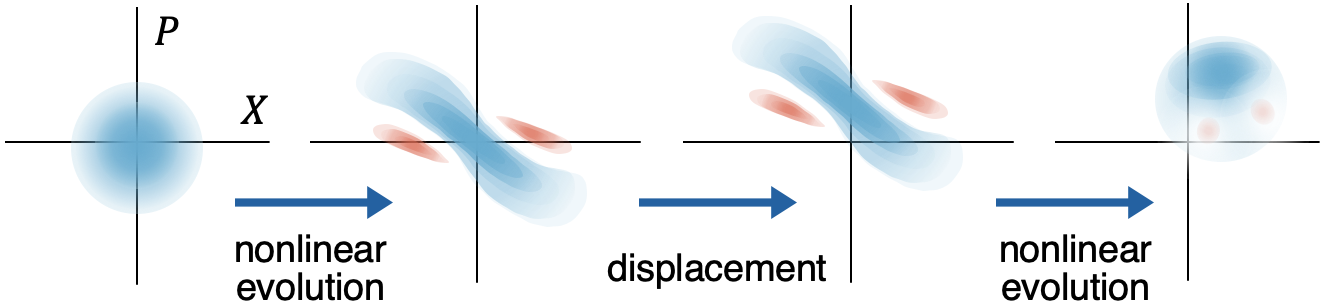}
	\end{center}
    \caption{\textbf{Illustration of the proposed metrological protocol.} Squeezing the ground state of a nonlinear oscillator results in a non-Gaussian state that might not have a quadrature with reduced quantum noise. Despite this, the state can still have high sensitivity to displacements, even if the information is hidden in high-order moments of quadrature operators. The method we propose to access the displacement amplitude consists instead in undergoing a second nonlinear evolution, which allows to keep the measurement linear.}
    \label{fig:0}
\end{figure}

Continuous variable (CV) quantum systems, such as optical fields or mechanical oscillators, constitute a platform of primary importance for quantum metrology applications. Examples include gravitational wave detection~\cite{LIGONP2013,LIGONPhysics2011,VirgoPRL2019,TsePRL2019}, force sensing~\cite{FabianNC2019,KatherineNature2019}, and the measurement of electric and magnetic fields~\cite{WangNC2019}. Largely explored in this context are squeezed states, namely Gaussian states that show along some phase-space quadrature an uncertainty that is below the quantum noise of the vacuum. Besides being relatively easy to be prepared experimentally, for example through a parametric process \cite{Wu86,Rugar91}, such states are also easy to be measured, as they can be fully characterised by linear quadrature (\ie homodyne) measurements.

One of the experimental factors limiting squeezing are the inevitable nonlinearities present in the system. In fact, highly squeezed states have large average number of excitations (\ie energy), as well as wavefunctions significantly distributed in phase space, which is manifested by the antisqueezed quadrature. When these go beyond the linear regime of the considered experimental platform, the state evolution become non-Gaussian, and squeezing gets degraded by a ``wrapping around'' of the state \cite{highorderSqPRApp,sq3dbPRXQ}. Similar results are found in the context of `crescent states', namely coherent states undergoing Kerr evolution \cite{GerryPRA1994,KitagawaPRA1986,KitagawaPRA1987,AndrewJOB2000}. This hinders metrological applications of the resulting state, despite the fact that non-Gaussian states can still have high sensitivity to perturbations~\cite{WojciechNature2001,JaewooPRL2011,SergeiNP2017,HyukjoonPRL2019,JunhuaPRL2018,VittorioNP2011}.

One of the main difficulties in doing quantum-enhanced metrology with non-Gaussian states lies in the fact that the parameter to be retrieved is encoded in high-order correlators~\cite{ManuelPRL2019}. This requires to access high moments of the measurement's probability distribution, or equivalently to perform non-Gaussian (\eg number-resolving) measurements, which is of challenging experimental implementation~\cite{HelmutScience2014,KaiPRL2022,LuckeScience2011,ThomasPRL2011,UweNP2022}. In addition, noise constraints for these observables become also very stringent. Therefore, nonlinearities are typically seen as limitations, and in experiments one tries to reduce them as much as possible.

A paradigmatic model where to study the interplay between squeezing and nonlinear interactions is given by the Hamiltonian of a Kerr oscillator with a squeezing drive
\begin{equation}\label{eq:Hdef}
    \hat{H} = \Delta \hat{a}^\dagger \hat{a} + \epsilon (\hat{a}^{\dagger2} + \hat{a}^2) - K \hat{a}^{\dagger2} \hat{a}^2 \;.
\end{equation}
Here $\Delta$ is the detuning between oscillator and drive, $\epsilon$ is the squeezing rate, and $K$ is the Kerr nonlinearity. Besides being interesting for the study of interesting processes such as chaotic dynamics \cite{MilburnPRA91}, tunneling \cite{WielingaPRA93},  coherent superpositions \cite{MiranowiczQO1990,GantsogQO1991}, phase transitions and blockade effects \cite{LeonskiPRA1994,ImamogluPRL1997}, Hamiltonian~\eqref{eq:Hdef} also attracted significant attention in the context of quantum information processing, since its ground state is a Schrödinger cat state that can be exploited for error-protected qubit encoding \cite{CochranePRA99,Goto16,Puri17}. This observation motivated the recent experimental implementation of Eq.~\eqref{eq:Hdef} for electromagnetic modes with superconductive devices~\cite{frattini2022squeezed,Venkatraman}, and in mechanical modes with acoustic resonators \cite{marti2023}.

Here we study the use of a squeezed Kerr oscillator for quantum metrology, and show that when sufficient control over the system's parameters is available, the presence of a nonlinearity can significantly improve the metrological performances even for simple quadrature measurements. The idea relies on preparing non-Gaussian states that, even when not showing reduced uncertainty compared to vacuum, can have high sensitivity to displacements. This can be then accessed by preceding the measurement step by an additional nonlinear evolution of the state, which we show to result in an effective measurement of higher-order moments of the quadratures. To conclude, we show that our results are robust to noise, and of immediate implementation in electrical and mechanical systems.

%\vspace{2mm}
\section{Squeezing limits} 
Let us begin with considering the task of preparing squeezed vacuum states with Eq.~\eqref{eq:Hdef}, and investigate the limitations posed by the nonlinear term. 

We imagine a protocol where a system that is initially in the ground state of the harmonic oscillator $\ket{0}$ evolves for $t\geq 0$ according to Eq.~\eqref{eq:Hdef}, due to the application of a parametric drive. During this dynamics, we are interested in studying the evolution of the state's minimum uncertainty quadrature, namely of $\Vmin(t) \equiv \min_{\theta} \text{Var}[(\hat{a} e^{-i\theta} + \hat{a}^\dagger e^{i\theta})/\sqrt{2}]$
, with $\text{Var}[\hat{A}]=\avg{\hat{A}^2}-\avg{\hat{A}}^2$ the variance of $A$. Since for coherent states $\Vmin=1/2$, observing $\Vmin<1/2$ implies a reduction of the quantum noise below the classical limit, and thus implies that the state is squeezed. 

\begin{figure}[t]
    \begin{center}
	\includegraphics[width=\columnwidth]{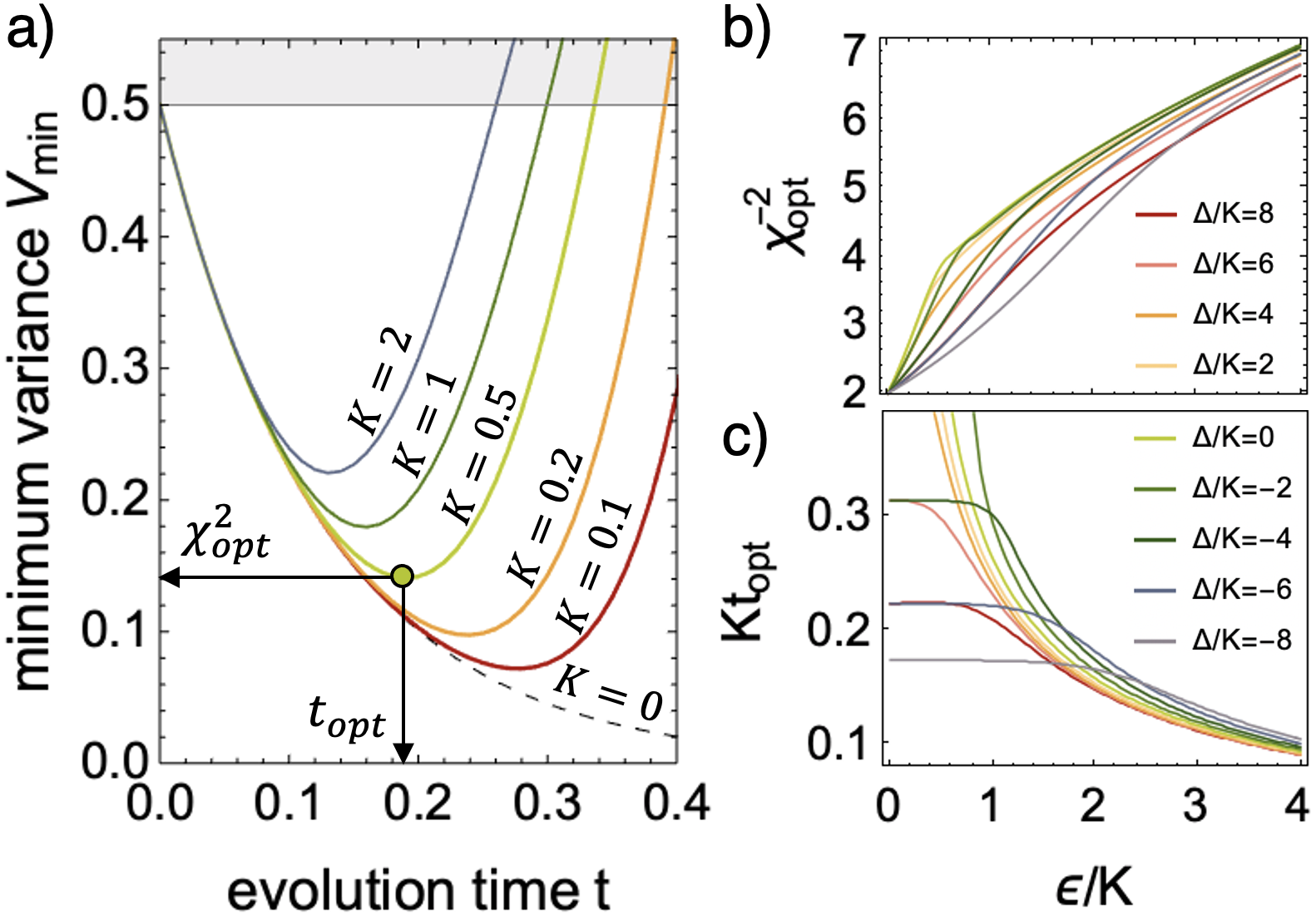}
	\end{center}
    \caption{\textbf{Optimal squeezing of a Kerr oscillator.} a) Minimum variance of the state $e^{-i\hat{H}t}\ket{0}$, fixing $\Delta=0$. For $K>0$ there is an optimal squeezing point, that we further investigate also as a function of $\Delta$. b) Squeezing level at the optimal point. c) Optimal squeezing time.}
    \label{fig1}
\end{figure}

For the dynamics we consider, there is in general no known analytic closed-form expression for $\Vmin$, which therefore has to be computed numerically.
We show in Fig.~\ref{fig1}a the plot of $\Vmin(t)$ for $(\Delta,\epsilon)=(0,2)$ and different values of $K$. For $K=0$ it is known that $\Vmin(t)=\frac{1}{2}e^{-4\epsilon t}$, meaning that an arbitrarily small uncertainty is achievable for sufficiently long times. For $K\neq 0$, however, we observe that $\Vmin$ attains a minimum value at a finite time $\topt$. This is expected, as the nonlinearity results in a non-Gaussian evolution of the state which limits the achievable squeezing \cite{highorderSqPRApp,sq3dbPRXQ}.
We thus define the parameter $\chi^{-2}_\text{opt}= 1/ \Vmin(\topt)$, which we will later show to be related to the state sensitivity, and show in Fig.~\ref{fig1}b,c the dependence of $\chi^{-2}_\text{opt}$ and $\topt$ on $\epsilon/K$, now also for different values of $\Delta$. Note that, higher squeezing can be prepared in a shorter time as $\epsilon/K$ increases, since the effect of the nonlinearity gets relatively less important.

Since squeezed states have a quadrature with reduced uncertainty, they can provide an advantage in metrological tasks~\cite{CavesPRD1981,MinPRL1987,GrangierPRL1987}. For this reason, it may look like as if the best strategy to achieve a larger advantage is to have $\epsilon/K$ as large as possible and stop the state preparation at $\topt$, since longer evolution times degrade $\Vmin$. However, as we will now show, this is not necessarily true.

%\vspace{2mm}
\section{Metrological advantage of non-Gaussian states} 
Let us remember that, in a typical quantum metrology scheme, the task is to estimate a parameter $d$ that is encoded in a probe state $\rho$ by a generator $\hat{G}$ as $\rho_d =e^{-id \hat{G}}\rho e^{id \hat{G}}$. A fundamental limit to the sensitivity is provided by the so-called quantum Cram$\acute{\text{e}}$r-Rao bound $\Delta^2 d \geq \Delta^2 d_{QCR} \equiv ( F_Q[\rho,\hat{G}])^{-1}$, where $F_Q[\rho,\hat{G}]$ is the quantum Fisher information (QFI). For a pure state the QFI is calculated from the variance of the generator as $F_Q[\rho,\hat{G}]=4\var[\hat{G}]$. Importantly, to achieve the maximum sensitivity $(\Delta^2 d_{QCR})^{-1}$ it is necessary to optimize the measurement that is performed on $\rho_d$ in order to estimate $d$. In general, if one measures $\op{M}$ then the achieved sensitivity is \cite{ManuelPRL2019}
\begin{equation}\label{Xi2Inv}
    \left( \Delta^2 d \right)^{-1} = \chi^{-2} [\rho,\hat{G},\hat{M}] \equiv  \frac{|\langle [\hat{G},\hat{M}] \rangle |^2}{\var[\hat{M}]} \;,
\end{equation}
which satisfies $\chi^{-2} [\rho,\hat{G},\hat{M}] \leq F_Q[\rho,\hat{G}]$~\cite{BraunsteinPRL1994}.

To now understand the connection between \eqref{Xi2Inv} and squeezing, let us consider the task of sensing the amplitude of a displacement from the measurement of a phase-space quadrature. We thus have $\hat{G}(\phi)=(\hat{a} e^{-i\phi} + \hat{a}^\dagger e^{i\phi})/\sqrt{2}$, the generator of a displacement along direction $\phi+\pi/2$, and $\hat{M}(\theta)=(\hat{a} e^{-i\theta} + \hat{a}^\dagger e^{i\theta})/\sqrt{2}$, the measurement along direction $\theta$. Looking at the numerator of Eq.~\eqref{Xi2Inv}, we note that the sensitivity is highest when $\hat{M}$ is perpendicular to $\hat{G}$, meaning when $\hat{M}$ is along the displacement direction, as we would expect. In this case we obtain $\chi^{-2}[\rho,\hat{G}(\theta+\pi/2),\hat{M}(\theta)] = 1/\var[(\hat{a} e^{-i\theta} + \hat{a}^\dagger e^{i\theta})/\sqrt{2}]$, and by further optimizing over the measurement direction $\theta$ we have $\chi^{-2} \equiv \max_{\phi,\theta}\chi^{-2} [\rho,\hat{G}(\phi),\hat{M}(\theta)] = 1/\Vmin$. 

These results shows us two things. First, for sensing displacements from quadrature measurements then $\Vmin$ (\ie the squeezing) is the correct figure of merit that needs to be optimized, but if the displacement is estimated from another type of measurement this might not be the case. Second, depending on the state $\rho$ we are considering then the choice of performing a quadrature measurement might not be the optimal one saturating the quantum Cram$\acute{\text{e}}$r-Rao bound, and thus not achieving $\chi^{-2} [\rho,\hat{G},\hat{M}]=F_Q[\rho,\hat{G}]$. This measurement choice is however proven to be optimal for sensing displacements with Gaussian states (see Sec. I, II and III of the SM~\cite{SM}). 

\begin{figure}[t]
    \begin{center}
	\includegraphics[width=85mm]{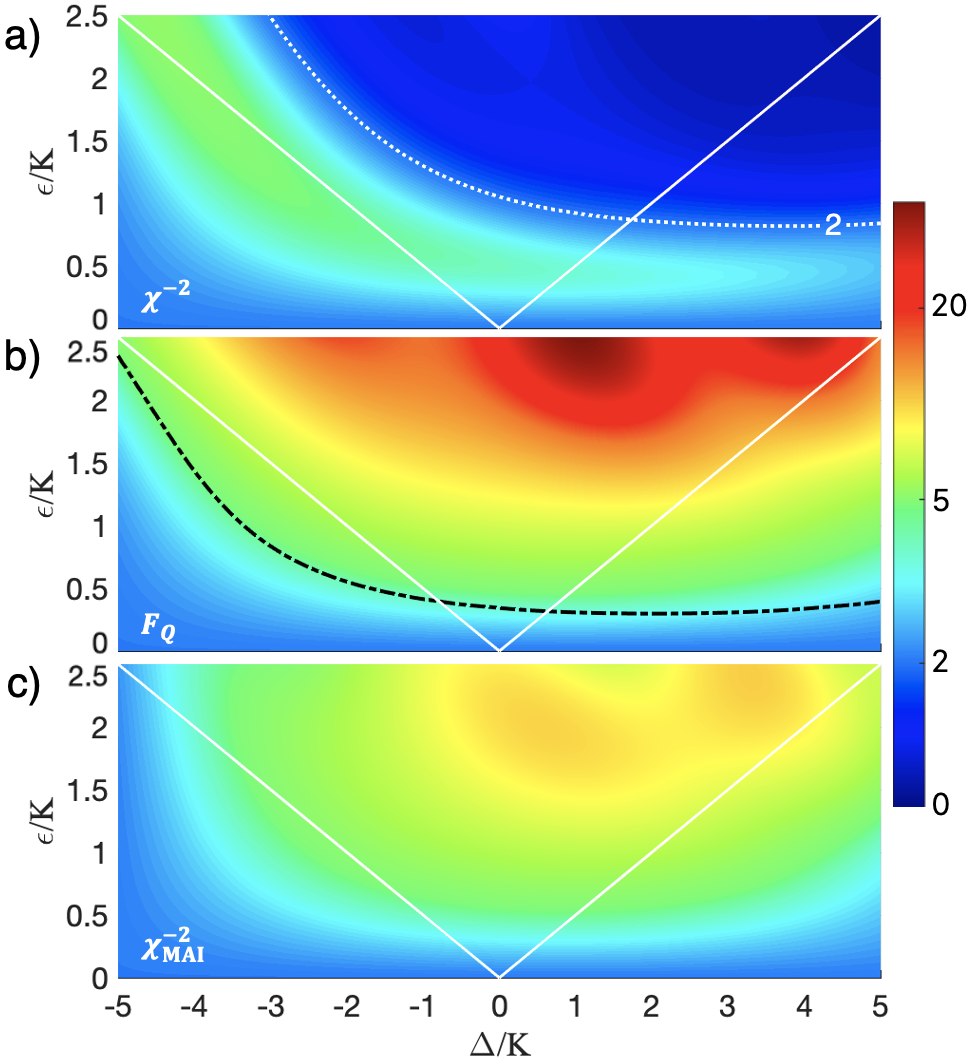}
	\end{center}
    \caption{\textbf{Sensitivity comparison.} For the state $e^{-i\hat{H}t}\ket{0}$ with $Kt=0.5$ we show:  a) squeezing $\chi^{-2}$, where the white dotted lines represents the standard-quantum-limit $\chi_{\text{SQL}}^{-2}=2$. b) QFI $F_Q$, where black dot-dashed line is a boundary $(F_Q-\chi^{-2})/F_Q=0.05$. c) Sensitivity for the MAI method $\chi_{\text{MAI}}^{-2}$. White solid lines are for $\epsilon=|\Delta/2|$, and indicate the classical phase diagram for a squeezed Kerr oscillator \cite{Venkatraman}. }
    \label{fig2}
\end{figure}

To illustrate this last point for the scenario introduced in the previous section, we compute squeezing and QFI of the states $\rho=e^{-i\hat{H}t}\ket{0}$ prepared through Eq.~\eqref{eq:Hdef} at $Kt=0.5$. We plot in Fig.~\ref{fig2}a,b the quantities $\chi^{-2}= 1/\Vmin$ and $F_Q\equiv \max_{\phi} F_Q[\rho,\hat{G}(\phi)]= \max_{\phi} 4\var[\hat{G}(\phi)]$, respectively, for different values of $\Delta/K$ and $\epsilon/K$. 
Figure~\ref{fig2}a shows that when linear quadrature measurements are performed, then a quantum-enhanced sensitivity, \ie $\chi^{-2}>2$, is attained only for a limited set of states. In general, one can also have $\chi^{-2}<2$, which indicates even worse sensitivity than the one achieved by coherent states (see also Fig.~\ref{fig1}a for $t>0.25$, when the coloured lines show $\Vmin>1/2$). When this is the case, since here we are dealing with pure states, it necessarily implies that the state is non-Gaussian. 
On the other hand, 
Fig.~\ref{fig2}b shows that any state (besides the trivial case $\epsilon/K=0$) has $F_Q >2$ and can thus show a quantum-enhanced sensitivity to displacements if the correct measurement is performed. 

Even if, strictly speaking, linear quadrature measurements are never optimal for $\epsilon/K>0$, there are regimes in which they are very close to being optimal. This is indicated by the region below the black line in Fig.~\ref{fig2}b, which corresponds to states for which $(F_Q-\chi^{-2})/F_Q \leq 0.05$, and thus to states that are faithfully approximated by being Gaussian. Outside this region, different measurements are required to approach the maximum sensitivity, which is what we want to investigate in the next sections by considering two strategies that are experimentally relevant.

%\vspace{2mm}
\section{Nonlinear measurements}
The first strategy consists of measuring higher order moments of phase-space quadratures, and it can be studies systematically in the following way~\cite{ManuelPRL2019}. We define $\textbf{M}^{(k)}$ to be a vector involving up to $k$th-order moments of the quadrature operators $\hat{X}=(\hat{a}+\hat{a}^\dagger)/\sqrt{2}$ and $\hat{P}=-i(\hat{a}-\hat{a}^\dagger)/\sqrt{2}$, such that, \eg linear quadrature measurements are described by $\textbf{M}^{(1)}=(\hat{X},\hat{P})$, while 2nd order one by $\textbf{M}^{(2)}=(\hat{X},\hat{P},\hat{X}^2,\hat{P}^2,(\hat{X}\hat{P}+\hat{P}\hat{X})/2)$. Then, we introduce the nonlinear squeezing parameter as~\cite{ManuelPRL2019}
\begin{equation}\label{Xi2InvNL}
    \chi_{(k)}^{-2} \equiv \max_{\phi,\vec{m}} \chi^{-2} [\rho, \hat{G}(\phi), \vec{m}\cdot \textbf{M}^{(k)}] \;.
\end{equation}
Crucially, this optimization task can be casted into an eigenvalue problem of easy solution (see Section I of the SM~\cite{SM}).
With higher-order moments involved, these parameters are capable of revealing quantum-enhanced sensitivites in a wider class of states, even beyond the Gaussian regime. Moreover, it holds the hierarchy $\chi_{(1)}^{-2}\leq \chi_{(2)}^{-2} \leq \cdots \leq F_Q$, showing that for sufficiently high $k$ one can attain the maximum sensitivity.

In squeezed Kerr oscillators the nonlinearity can results in highly non-Gaussian states, whose metrological advantage is unlocked only for measurements of sufficient high order. Observing $\chi_{(1)}^{-2} = \chi^{-2} <2$ implies that linear quadrature measurements are not sufficient, and thus that $k>1$ is necessary. Based on this observation, we computed $\chi_{(2)}^{-2}$ but, perhaps surprisingly, did not find any advantage compared to $\chi_{(1)}^{-2}$. A careful exploration of the terms involved shows that this is due to the fact that commutators between the generator and second moments of the quadrature give terms linear in the quadrature, whose expectation value is zero for the states we consider (see SM~\cite{SM}, Sec. I). This means that, in our scenario, considering $\textbf{M}^{(2)}$ does not provide any advantage compared to $\textbf{M}^{(1)}$. In order to see an advantage one would need to consider at least $\textbf{M}^{(3)}$, which requires a massive increase in the measurement statistics and low detection noise. This can be of impractical implementation in several experimental situations, and it is thus viable to consider also alternative strategies.

%\vspace{2mm}
\section{Measurement-After-Interaction (MAI) protocol} 
The second approach we consider consists of preceding a linear quadrature measurement by a time-reversed evolution with Eq.~\eqref{eq:Hdef}, \ie $e^{+i\hat{H}t}$. A similar idea has been investigated for spin states both theoretically \cite{EmilyPRL2016,FlorianPRL2016,TommasoPRA2016,YoucefPRL2021,YoucefCRP2022,SamuelPRL2017} and experimentally \cite{HostenScience2016,SimoneNP2022}. For CV systems, an experiment with Gaussian states and transformations has been presented in Ref.~\cite{BurdScience2019}. In this framework, reversing the evolution results in an amplification of the signal to be detected, but also of the quantum noise (see Fig.~1 of Refs.~\cite{FlorianPRL2016,BurdScience2019}). For this reason, an advantage is obtained only in the presence of detection noise limiting the measurement resolution, as the ratio between the signal and the quantum noise level remains constant (see Sec. IV and V of the SM~\cite{SM}). In fact, linear quadrature measurements are already optimal for sensing displacements with Gaussian states, in the sense that they achieve $\chi^{-2}=F_Q$ (see Sec. III of the SM~\cite{SM}). 

In the following we will show that the MAI protocol can provide tremendous sensitivity enhancements when considering CV non-Gaussian states and transformations. Necessary conditions for this protocol to be viable are: i) the capability of implementing a time-reversed evolution and ii) low enough noise to tolerate a second time evolution of the state. For point i), it is sufficient to be able to invert the sign of the parameters in the Hamiltonian. In the specific case of Eq.~\eqref{eq:Hdef}, $\Delta$ and $\epsilon$ are easily tuned by the parametric drives, while $K$ can be tuned by changing the anharmonicity of a trapping potential (\eg in the case of a trapped ion) or the coupling between the bosonic mode and a two-level system (\eg in a circuit-QED setup).

\begin{figure}[t]
    \begin{center}
	\includegraphics[width=85mm]{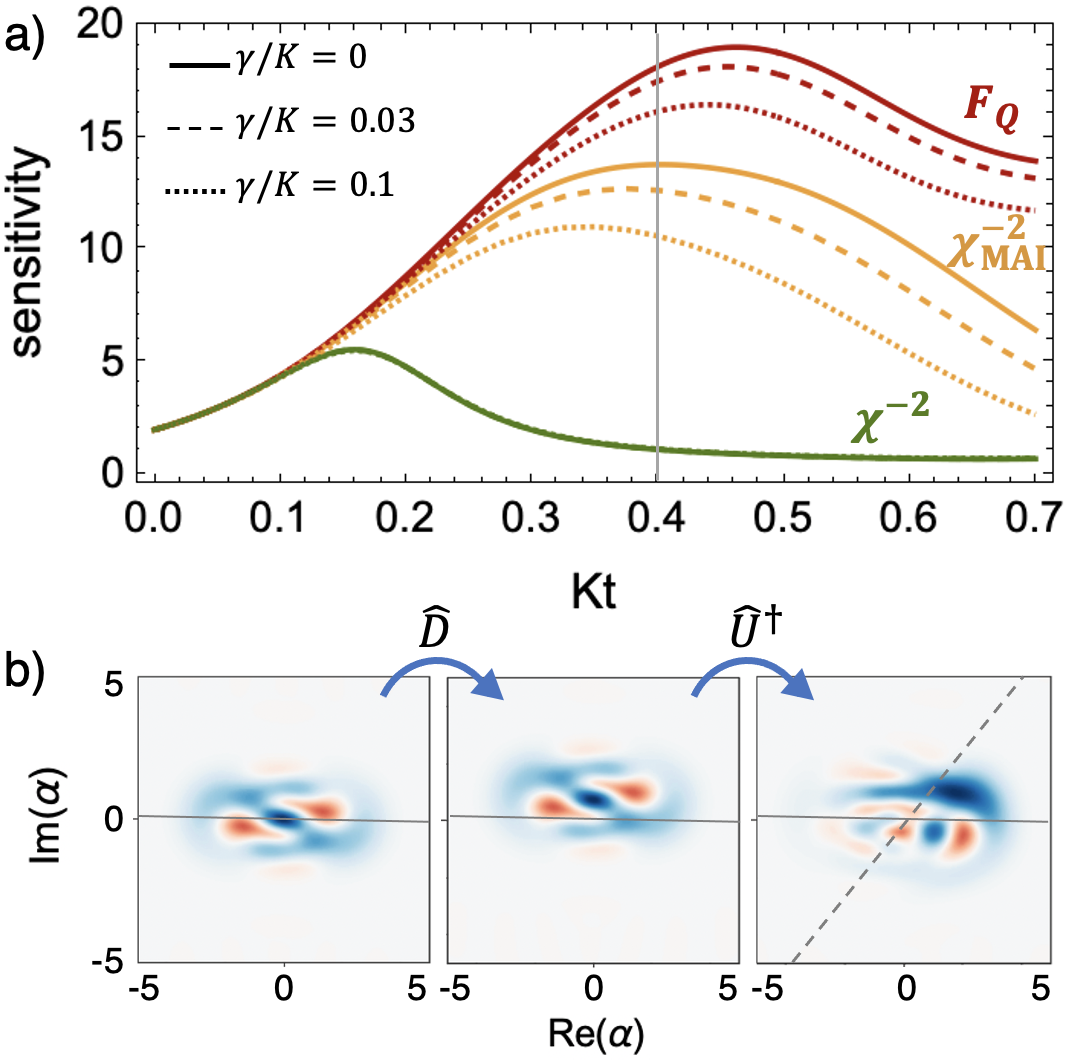}
	\end{center}
    \caption{\textbf{Noise robustness.} a) sensitivities $F_Q, \chi_{\text{MAI}}^{-2}, \chi^{-2}$ obtained for $\Delta/K=0, \epsilon/K=2$, and different levels of energy relaxation rates $\gamma/K$. b) Wigner functions at different steps of the MAI protocol for $Kt=0.4$ and $\gamma/K=0.1$. From left to right: prepared state, displacement, time-reversed nonlinear evolution. Solid and dashed lines are optimal directions of the generator and linear quadrature measurement, respectively.}
    \label{fig3}
\end{figure}

Formally, the additional evolution that precedes the measurement can be absorbed into a redefinition of the measurement operator. In our case we have $\hat{M}_{\text{MAI}}(\theta)=\hat{U} (a e^{-i\theta} + \ad e^{i\theta}) \hat{U}^\dagger /\sqrt{2}$, where $\hat{U}=e^{-iHt}$, from which we define
\begin{equation}\label{Xi2InvMAI}
    \chi_{\text{MAI}}^{-2} \equiv \max_{\phi,\theta} \chi^{-2} [\rho, \hat{G}(\phi), \hat{M}_{\text{MAI}}(\theta)] \;.
\end{equation}
We plot this parameter in Fig.~\eqref{fig2}c for the states $e^{-iHt}\ket{0}$ prepared at $Kt=0.5$. From a comparison with the $F_Q$ shown in Fig.~\eqref{fig2}b we are able to conclude that the MAI protocol can attain a sensitivity close to optimal. In particular, this is true also for the non-Gaussian regime, where by looking at Fig.~\eqref{fig2}a we see that by performing linear quadrature  without the time-reversed dynamics one would only get $\chi^{-2}<2$. A more quantitative comparison will be discussed later in Fig.~\ref{fig3}, while an analysis of the scaling with $N$ can be found in Section VI of the SM~\cite{SM}.

%\vspace{2mm}
\section{Robustness to losses}
To show that the MAI protocol gives an actual advantage in realistic scenarios we have to consider the effect of experimental imperfections. 
During both state preparation and time-reversal dynamics the evolution of the system is affected by inevitable losses and decoherence, which ultimately limit the maximum duration of the protocol. If these are too severe compared to the robustness of the MAI protocol, then no advantage is obtained. 

To study this in detail, in our numerical simulations we replace the unitary time evolutions $\hat{U}$ and $\hat{U}^\dagger$ by evolutions according to a Master equation. We focus on losses (\ie energy relaxation), which pose a major limitation in CV systems. These can be described by a jump operator $\sqrt{\gamma}\hat{a}$, where $\gamma=1/T_1$ is the energy relaxation rate.

Since here we are dealing with mixed states and non-unitary evolutions, calculating the sensitivity becomes more tedious.
The QFI for a general state $\rho$ is computed as $F_Q[\rho,\hat{G}]=\sum_{kl}\frac{(\lambda_k-\lambda_l)^2}{(\lambda_k+\lambda_l)} |\bra{k}\hat{G}\ket{l}|^2$, where $\lambda_k$ and $\ket{k}$ are the eigenvalues and eigenstates of $\rho$, respectively. To calculate $\chi^{-2}$ and $\chi^{-2}_\text{MAI}$ we use the fact that $|\langle [\hat{G},\hat{M}] \rangle |^2=\big{|}\frac{\partial}{\partial d} \text{Tr}[\hat{M} \rho_d] \big{|}^2_{d=0}$, where $\rho_d =e^{-id \hat{G}}\rho e^{id \hat{G}}$, and then discretize the derivative numerically by applying a small displacement to the state.

We show a comparison of the sensitivities in Fig.~\ref{fig3}a, for $\Delta/K=0$, $\epsilon/K=2$, and different loss rates $\gamma/K$ (even stronger than the one observed in experiments \cite{Venkatraman}). The sensitivity $\chi^{-2}$ obtained from linear quadrature measurements is almost unaffected by losses, but as we have seen it approaches $F_Q$ only for small times ($Kt\approx 0.1$). On the other hand, the sensitivity $\chi^{-2}_\text{MAI}$ obtained from the MAI protocol is always significantly larger than $\chi^{-2}$, and it approaches $F_Q$ for a longer time interval ($Kt\approx 0.25$). We thus conclude that the MAI protocol is robust, in the sense that a large amount of losses is required before having the maximum of $\chi^{-2}_\text{MAI}$ smaller than the maximum of $\chi^{-2}$ (see also Section VII of the SM \cite{SM}). Let us emphasize that the observed performance of the MAI protocol is remarkable, considering that doubling the time evolution significantly increases losses. 

To have a better understanding of the MAI protocol, we plot in Fig.~\ref{fig3}b the Wigner function of the state at different steps. First, a non-Gaussian state is prepared by evolving $\ket{0}$ according to Eq.~\eqref{eq:Hdef} with $\Delta/K=0$, $\epsilon/K=2$ and $Kt=0.4$. Then, the state is subject to the displacement we want to sense, followed by an evolution with $-\hat{H}$ for another $Kt=0.4$. Finally, a linear quadrature measurement is performed on the state to estimate the displacement amplitude. Solid and dashed lines in Fig.~\ref{fig3}b indicate the optimal generator and measurement directions, respectively.

%\vspace{2mm}
\section{Conclusions} 
We addressed the problem of optimally using a squeezed Kerr oscillator for a metrological task. In particular, we focus on sensing displacements with the non-Gaussian states that result from the nonlinear evolution of this system, and investigate measurement strategies to approach the highest sensitivity. We show that, while a direct quadrature measurement requires to access high-order moments, preceding the measurement by a time-reversed nonlinear evolution allows to achieve high sensitivities even for first moments. Crucially, the protocol we propose is robust to noise, and can be implemented in current experiments for quantum-enhanced sensing of \eg electromagnetic fields \cite{frattini2022squeezed,Venkatraman} or forces \cite{marti2023}. Future works could address the interesting problem of multiparameter estimation for entangled nonlinear oscillators \cite{WieslawJOB2004,OlsenPRA2006,SaidJPB2006,JoannaQIP2017}.

\vspace{2mm}
\textit{Acknowledgments.--} We thank S. Liu, F. Sun and M. Tian for helpful discussions. This work is supported by the National Natural Science Foundation of China (Grants No.~11975026, No.~12125402, and No.~12147148), and the Innovation Program for Quantum Science and Technology (Grant No. 2021ZD0301500). MF was supported by the Swiss National Science Foundation Ambizione Grant No. 208886, and The Branco Weiss Fellowship -- Society in Science, administered by the ETH Z\"{u}rich.

\bibliographystyle{apsrev4-1} % Tell bibtex which bibliography style to use
\bibliography{mylib3.bib}

\clearpage
\newpage

\begin{widetext}

\section*{Supplemental material for ``Quantum metrology with a squeezed Kerr oscillator''}

\section{Metrological sensitivity}
In a metrological task, a parameter $d$ is encoded in a probe state $\rho$ by a generator $\hat{G}$ via $\rho_d = e^{-d\hat{G}}\rho e^{id\hat{G}}$. The goal is then to estimate the parameter $d$ with the smallest uncertainty $\Delta d$ from the measurement of an observable $\hat{M}$ on $\rho_d$. The uncertainty in the estimation can be expressed as
\begin{align}\label{eq:Deltad2}
\Delta^2 d  =\frac{ \var [\hat{M}] }{|\partial_d \langle \hat{M} \rangle |^2} \Bigg{|}_{d=0} =\frac{\var[\hat{M}]}{ |\langle[\hat{G},\hat{M}]\rangle|^2 } \;.
\end{align}
In our work, we consider a continuous variable scenario where the parameter to be estimated is the amplitude $d=\sqrt{2}|\alpha|$ of a displacement $\hat{D}(\alpha)=e^{\alpha\hat{a}^\dagger-\alpha^*\hat{a}}$. 

It is convenient to expand $\hat{G}$ and $\hat{M}$ as $\hat{G}=\vec{n}^T\cdot \textbf{G}$ and $\hat{M}=\vec{m}^T\cdot\textbf{M}$, where $\textbf{G},\textbf{M}$ are vectors of observables. The sensitivity, \ie the inverse of Eq.~(\ref{eq:Deltad2}), can now be written as
\begin{align}\label{eq:sensDef}
\chi^{-2}[\rho,\hat{G},\hat{M}] \equiv \frac{|\langle[\hat{G},\hat{M}]\rangle|^2 }{\var[\hat{M}]} = \frac{|\vec{n}^T\textbf{C}[\rho,\textbf{G},\textbf{M}]\vec{m} |^2}{ \vec{m}^T \boldsymbol{\Gamma}[\rho,\textbf{M}] \vec{m} } \;,
\end{align}
where $\textbf{C}[\rho,\textbf{G},\textbf{M}]$ is the commutator matrix with elements $\left(\textbf{C}[\rho,\textbf{G},\textbf{M}]\right)_{ij}=-i\langle [\hat{G}_i,\hat{M}_j]\rangle_\rho$, and $\boldsymbol{\Gamma}[\rho,\textbf{M}]$ is the covariance matrix with elements $\left( \boldsymbol{\Gamma}[\rho,\textbf{M}] \right)_{ij}=\text{Cov}\left[\hat{M}_i,\hat{M_j} \right]_\rho$. For a given probe state $\rho$, the maximum sensitivity can be obtained by optimizing Eq.~\eqref{eq:sensDef} over both vectors $\vec{n}$ and $\vec{m}$, which has been proved to be equivalent as the maximum eigenvalue of the moment matrix $\mathcal{M}\equiv\textbf{C}\boldsymbol{\Gamma}^{-1}\textbf{C}^T$ \cite{ManuelPRL2019}. In symbols, the maximum sensitivity is thus
\begin{align}\label{eq:supp_chiMax}
\chi^{-2} \equiv \max_{\vec{m},\vec{n}}\chi^{-2}[\rho,\hat{G},\hat{M}] =\lambda_{\text{max}} (\mathcal{M}) \;,
\end{align}
where the optimal direction $\vec{n}_{\text{opt}}$ for the generator is the eigenvector corresponding to the maximum eigenvalue $\lambda_{\text{max}}$, and the optimal vector for the measurement operator is $\vec{m}_{\text{opt}}=\alpha \boldsymbol{\Gamma}^{-1}\textbf{C}^T\vec{n}_{\text{opt}} $, with $\alpha$ a normalization constant.

If only linear quadratures measurements are taken into account, the vectors of operators are $\textbf{G}^{(1)}=\textbf{M}^{(1)}=\{\hat{X},\hat{P}\}$. The generator and the measurement can thus be expressed in terms of two angles, $\phi$ and $\theta$, namely
\begin{align}
\hat{G}(\phi) = \left( \hat{a} e^{-i\phi}+\hat{a}^\dagger e^{i\phi} \right)/\sqrt{2} \;, \\
\hat{M}(\theta) = \left( \hat{a} e^{-i\theta}+\hat{a}^\dagger e^{i\theta} \right)/\sqrt{2} \;.
\end{align}
For this choice, the maximum sensitivity is obtained when $\vec{m}\perp \vec{n}$, so that it can be simplified as $\chi_{(1)}^{-2}= (\min_{\theta}\text{Var}[\hat{M}(\theta)] )^{-1}$.

One way to possibly increase the sensitivity is to include in the vector of observables $\textbf{M}$ higher-order measurement operators. Let us denote with $\textbf{M}^{(k)}$ the vector including at most $k$-th order operators. For example, the second-order $\textbf{M}^{(2)}$ is 
\begin{align}
\textbf{M}^{(2)} = \{\textbf{M}^{(1)},\hat{X}^2,\hat{P}^2, (\hat{X}\hat{P}+\hat{P}\hat{X})/2 \} \;,
\end{align}
and the third-order $\textbf{M}^{(3)}$ is 
\begin{align}
\textbf{M}^{(3)} = \{ \textbf{M}^{(2)},\hat{X}^3,\hat{P}^3,(\hat{X}\hat{P}\hat{P}+\hat{P}\hat{X}\hat{P}+\hat{P}\hat{P}\hat{X})/3, (\hat{P}\hat{X}\hat{X}+\hat{X}\hat{P}\hat{X}+\hat{X}\hat{X}\hat{P})/3 \} \;.
\end{align}
Note here that, since the generator $\hat{G}$ is associated with a displacement in phase-space, the vector $\textbf{G}=\textbf{G}^{(1)}$ is always linear in the quadratures. Combining these object, the sensitivity is
\begin{align} 
\chi_{(k)}^{-2} \equiv \max_{\vec{m},\vec{n}}  \frac{|\langle[\hat{G},\hat{M}^{(k)}]\rangle|^2 }{\var[\hat{M}^{(k)}]} =\lambda_{\text{max}} \left( \mathcal{M}^{(k)} \right) \;,
\end{align}
where $\mathcal{M}^{(k)}=\textbf{C}[\rho,\textbf{G},\textbf{M}^{(k)}]\boldsymbol{\Gamma}^{-1}[\rho,\textbf{G},\textbf{M}^{(k)}]\textbf{C}^T[\rho,\textbf{G},\textbf{M}^{(k)}]$ is the second-order moment matrix.

For a state $|\psi\rangle=e^{-i\hat{H}t}|0\rangle$, with $H$ the squeezed Kerr oscillator Hamiltonian considered in the main text, the expectation value of a linear quadrature measurement along any directions is zero, \eg $\langle \hat{X} \rangle =\langle \hat{P} \rangle=0$. Because of this, if we now calculate the second-order sensitivity $\chi_{(2)}^{-2}$, we will find that the commutator between the generator $\hat{G}(\phi)$ and any second-order measurement operators is zero. The resulting commutator matrix is thus
\begin{align}
\textbf{C}[\rho,\textbf{G},\textbf{M}^{(2)}]= -i \left( 
\begin{matrix}
0 & 1 & 0 & 0 & 0 \\
-1 & 0 & 0 & 0 & 0
\end{matrix}
\right) \;,
\end{align}
meaning that the moment matrix will be $\mathcal{M}^{(2)}=\mathcal{M}^{(1)}$.
This result leads to the conclusion that, for the case we considered, second-order measurements are insufficient to provide an advantage over linear measurements, \ie $\chi_{(2)}^{-2}=\chi_{(1)}^{-2}$. Hence, to see an advantage, at least third-order measurements are required, which brings experimental challenges for the massive measurement data and low detection noise requirements.

\section{Quantum Fisher information}
The quantum Fisher information (QFI) $F_Q[\rho,\hat{G}]$ quantifies the sensitivity of a probe state $\rho$ with respect to a perturbation generated by $\hat{G}$. It can be expressed as \cite{BraunsteinPRL1994}
\begin{align}\label{eq:QFIgeneral}
F_Q[\rho,\hat{G}]=\sum_{\substack{k,l\\ \text{s.t. } \lambda_k+\lambda_l>0 }}\frac{(\lambda_k-\lambda_l)^2}{(\lambda_k+\lambda_l)} |\bra{k}\hat{G}\ket{l}|^2 \;,
\end{align}
where $\lambda_{k}$ are the eigenvalues of $\rho$, and $\ket{k}$ their associated eigenvectors. For pure states $\rho=|\psi\rangle\langle \psi|$, Eq.~\eqref{eq:QFIgeneral} is simply $F_Q[|\psi\rangle,\hat{G}]=4\var[\hat{G}]_{|\psi\rangle}$, meaning that the QFI is proportional to the variance of the generator.
In our work, the perturbations we consider are displacements of the state, meaning that the associated generator is linear in the quadratures $\hat{G}=\vec{n}\cdot\textbf{G}^{(1)}$.
The maximum QFI is attained for displacements along the state's most sensitive direction, and is $F_Q=\max_{\vec{n}} F_Q[\rho,\hat{G}]$.
This maximization over $\vec{n}$ can be efficiently performed by computing the maximum eigenvalue of the $2\times 2$ matrix with elements
\begin{align}
     [\mathbf{F}_Q]_{ij} = 2 \sum_{\substack{k,l\\ \text{s.t. } \lambda_k+\lambda_l>0 }} \frac{(\lambda_k-\lambda_l)^2}{(\lambda_k+\lambda_l)} \langle k|\hat{G}_i|l \rangle \langle l | \hat{G}_j |k\rangle \;,
\end{align}
where $\hat{G}_{i}\in\{\hat{X},\hat{P}\}$, so that we have $F_Q=\lambda_{\text{max}} (\mathbf{F}_Q)$. We used this result to efficiently compute the maximum QFI for mixed states, like for Fig.~4(a) in the main text.

\vspace{2mm}

Eq.~\eqref{eq:QFIgeneral} can be computed analytically for Gaussian states, giving as result \cite{Serafini2017}
\begin{align}\label{eq:QFIgauss}
    F_Q[\rho_d]=\frac{1}{2(1+\mathcal{P}(\rho_d)^2)} \text{Tr}\left[(\bm{\Gamma}^{-1}[\rho_d]\frac{\partial}{\partial d}\bm{\Gamma}[\rho_d])^2)\right]
+\left(\frac{\partial\avg{\mathbf{\hat{r}}}_{\rho_d}}{\partial d}\right)^T\mathbf{\Gamma}^{-1}[\rho_d]\left(\frac{\partial\avg{\mathbf{\hat{r}}}_{\rho_d}}{\partial d}\right) \;,
\end{align}
where $\mathcal{P}(\rho)=\text{Tr}[\rho^2]=(2\sqrt{\det\mathbf{\Gamma}[\rho]})^{-1}$ is the state purity, $\avg{\mathbf{\hat{r}}}_{\rho_d}=(\avg{\hat{X}}_{\rho_d},\avg{\hat{P}}_{\rho_d})^T$ is the vector of first moments, and $\bm{\Gamma}^{-1}[\rho_d]$ is the covariance matrix.
In the most general case, a Gaussian state can be expressed as a displaced squeezed thermal state, that is
\begin{equation}\label{eq:rhoG}
    \rho(\alpha,\xi,\overline{n}_T) = \hat{D}(\alpha)\hat{S}(\xi)\rho_{T} \hat{S}^{\dagger}(\xi)\hat{D}^{\dagger}(\alpha) \;,
\end{equation}
where $\rho_{T}$ is a thermal state with average number of thermal excitations $\overline{n}_T$, $\hat{D}(\alpha)$ is the displacement operator, and $\hat{S}(\xi) \equiv e^{(\xi^* \hat{a}^2 - \xi\hat{a}^{\dagger2})/2}$ is the squeezing operator with $\xi = r e^{i\zeta}$ . Gaussianity implies that these states can be fully described by first and second moments of quadrature operators. For this reason, a Gaussian state is fully determined by its displacement vector
\begin{equation}
\avg{\mathbf{\hat{r}}}_{\rho(\alpha,\xi,\overline{n}_T)}=\begin{pmatrix}
        \sqrt{2}\Re[\alpha]\\
        \sqrt{2}\Im[\alpha]
    \end{pmatrix} \;,
\end{equation}
and the covariance matrix
\begin{equation}\label{eq:covGauss}
    \Gamma[\rho(\alpha,\xi,\overline{n}_T)] = \dfrac{(1 + 2 \overline{n}_T)}{2} \begin{pmatrix}
\cosh(2r)-\sinh(2r)\cos\zeta & -\sinh(2r)\sin\zeta \\
-\sinh(2r)\sin\zeta & \cosh(2r)+\sinh(2r)\cos\zeta
\end{pmatrix} \;.
\end{equation}

For the case of phase-space displacements, $\partial_d \bm{\Gamma}[\rho_d] =0$ because the covariance matrix is invariant under translations. Eq.~\eqref{eq:QFIgauss}
\begin{align}\label{eq:QFIgaussDisp}
    F_Q[\rho,\op{G}(\phi)] &= \left(\frac{\partial\avg{\mathbf{\hat{r}}}_{\rho_d}}{\partial d}\right)^T\mathbf{\Gamma}^{-1}[\rho_d]\left(\frac{\partial\avg{\mathbf{\hat{r}}}_{\rho_d}}{\partial d}\right) \notag \\
    &=\frac{\vec{n}^T\mathbf{\Gamma}[\rho_d]\vec{n}}{\det\mathbf{\Gamma}[\rho_d]} \;,
\end{align}
where $\vec{n}=(\cos\phi,\sin\phi)^T$, and $\det\mathbf{\Gamma}[\rho]=\var[\op{X}]\var[\op{P}]-\cov[\op{X},\op{P}]^2$. In the second line, we used $\bm {A} ^{-1}={\frac {1}{\det \bm {A}}}\Omega^T\bm{A}^T\Omega$ for a $2\times 2$ matrix $\mathbf{A}$ and $(\partial/\partial d) \langle\hat{\bm{r}}\rangle_{\rho_d}=(-\sin\phi,\cos\phi)^T=\Omega^T\vec{n}$, where $\Omega=\bigl(\begin{smallmatrix}
 0 & 1\\ -1 & 0
\end{smallmatrix}\bigr)$ is the symplectic form. Plugging in Eq.~\eqref{eq:QFIgaussDisp} the covariance matrix Eq.~\eqref{eq:covGauss} we finally obtain that the QFI of a Gaussian state for displacements is
\begin{equation}\label{eq:QFIgaussDispExp}
    F_Q[\rho,\op{G}(\phi)] = \dfrac{2 \left( \cosh (2 r)-\sinh (2 r) \cos (\zeta -2 \phi ) \right)}{(1 + 2 n_T)} \;.
\end{equation}
The maximum of this expression is obtained for $\phi=\zeta/2-\pi/2$, and is
\begin{equation}\label{eq:maxQFIgauss}
    F_Q = \dfrac{2 e^{2r}}{(1+2\overline{n}_T)} \;.  
\end{equation}

If we consider a squeezed vacuum state (\ie $\alpha=0$ and $\overline{n}_T=0$), we can use the relation $N\equiv\avg{\op{a}^\dagger\op{a}}=\sinh^2 r$ to derive how the QFI scales with the average number of particles. This is
\begin{equation}
    F_Q = 2\left(1+2N+2\sqrt{N(N+1)} \right) \simeq (4 + 8 N) \quad\text{for}\;N\rightarrow\infty \;.
    \label{eq:scalingSqQFI}
\end{equation}

\section{Optimality of linear quadrature measurements}

We might ask whether linear measurements are optimal for detecting displacements using Gaussian states, in the sense that they allow to saturate the Cramér-Rao bound $\chi^{-2}=F_Q$.

In Eq.~\eqref{eq:QFIgaussDispExp} we have computed the QFI for Gaussian states under displacements. We now proceed with calculating $\chi^{-2}$ as defined in Eq.~\eqref{eq:supp_chiMax}.

First, considering a generator $\hat{G}=\vec{n}^T\cdot \{\op{X},\op{P}\}$ with $\vec{n}=(\cos\phi,\sin\phi)^T$ and a measurement $\hat{M}=\vec{m}^T\cdot \{\op{X},\op{P}\}$ with $\vec{m}=(\cos\theta,\sin\theta)^T$ , we have $|\langle [\hat{G},\hat{M}] \rangle|^2 = \sin^2(\theta-\phi)$. Then, the variance in the denominator of Eq.~\eqref{eq:sensDef} is simply $\var[\hat{M}]=\vec{m}^T \boldsymbol{\Gamma}[\rho,\textbf{M}] \vec{m}$, with the covariance matrix given by Eq.~\eqref{eq:covGauss}. Combining these results imply that the linear sensitivity to displacements of a Gaussian state is
\begin{align}
\chi^{-2}[\rho,\hat{G}(\phi),\hat{M}(\epsilon)] = \dfrac{2 \sin^2(\theta-\phi)}{(1+2 \overline{n}_T)\left(\cosh(2r)-\cos(\zeta-2\theta)\sinh(2r) \right)} \;.
\end{align}
It is now easy to check that by choosing $\theta=\zeta/2$, such that the denominator is minimized, and $\phi=\zeta/2-\pi/2$, such that the numerator is maximized, we obtain for Eq.~\eqref{eq:supp_chiMax}
\begin{align}
\chi^{-2} = \dfrac{2 e^{2 r}}{(1+2 \overline{n}_T)} \;.
\end{align}
This result coincides with Eq.~\eqref{eq:maxQFIgauss}, which allows us to conclude that linear measurements are optimal for sensing displacements with Gaussian states.

\section{MAI method for Gaussian states} 

In the scenario of displacement sensing using Gaussian states, Eq.~\eqref{eq:QFIgaussDispExp} shows us that the sensitivity be increased by the action of the squeezing operator
\begin{align}
\hat{S} = e^{(\xi^* \hat{a}^2 - \xi\hat{a}^{\dagger2})/2} \;,
\end{align}
where $\xi=r e^{i\zeta}$ and $r$ determines squeezing extent and $\zeta$ defines the squeezing direction. The action of this operator is
\begin{align}
\hat{S}^\dagger \hat{a} \hat{S} &=\cosh(r)\hat{a} -e^{i\zeta} \sinh(r) \hat{a}^\dagger \;, \\
\hat{S}^\dagger \hat{a}^\dagger \hat{S} &=\cosh(r)\hat{a}^\dagger -e^{-i\zeta} \sinh(r) \hat{a} \;.
\end{align}

For Gaussian states, the anti-squeezing and squeezing directions corresponds to the optimal generator $\hat{G}_{\text{opt}}$ and the optimal measurement $\hat{M}_{\text{opt}}$, respectively. As these two directions are orthogonal, we have $\vert[\hat{G}_{\text{opt}},\hat{M}_{\text{opt}}]\vert=1$, which will be useful later. 
Note that, upon action of the squeezing operator, these two directions transform as
\begin{align}
\hat{S}^\dagger \hat{G}_{\text{opt}} \hat{S} &= e^r \hat{G}_{\text{opt}} \;, \\
\hat{S}^\dagger \hat{M}_{\text{opt}} \hat{S} &= e^{-r} \hat{M}_{\text{opt}} \;.
\end{align}

Let us consider now a pure Gaussian state $|\psi_S\rangle=\hat{D}\hat{S}|0\rangle$, but keeping in mind that the following discussion can easily be generalized to mixed Gaussian states. Since the action of the displacement operators $\hat{D}(\alpha)$ gives no contribution to the sensitivity, we can consider only the case of squeezed vacuum states $\hat{S}|0\rangle$, whose optimal sensitivity to displacements is 
\begin{align}
\chi^{-2} &= \frac{|\langle [\hat{G}_{\text{opt}},\hat{M}_{\text{opt}}] \rangle|^2}{\var[\hat{M}_{\text{opt}}]} \nonumber \\
&= \frac{1}{\langle 0 |\hat{S}^\dagger \hat{M}_{\text{opt}}^2\hat{S}|0\rangle -\langle 0 |\hat{S}^\dagger \hat{M}_{\text{opt}} \hat{S} |0 \rangle} \nonumber \\
&= \frac{1}{\langle 0 | \left( \hat{S}^\dagger \hat{M}_{\text{opt}} \hat{S} \right) \left( \hat{S}^\dagger \hat{M}_{\text{opt}} \hat{S} \right) |0 \rangle  -\langle 0 |\hat{S}^\dagger \hat{M}_{\text{opt}} \hat{S} |0 \rangle^2 } \nonumber \\
&= \frac{1}{ e^{-2r} \left(\langle 0 | \hat{M}_{\text{opt}}^2 | 0 \rangle -\langle 0 |\hat{M}_{\text{opt}} |0 \rangle^2 \right) } \nonumber \\
&= 2e^{2r} \;. \label{eq:sensGLinM}
\end{align}

We are now interested in comparing the sensitivity Eq.~\eqref{eq:sensGLinM} obtained by performing linear quadrature measurements, with the one achieved by a MAI protocol.
In a MAI protocol, the optimal generator is still the same $\hat{G}_{\text{opt}}$ as before, but the optimal measurement operator $\hat{M}_{\text{MAI}}$ takes the form
\begin{align}
\hat{M}_{\text{MAI}} &= \hat{S} \hat{M}_{\text{opt}} \hat{S}^\dagger \;.
\end{align}
The sensitivity of the pure Gaussian state $|\psi_S\rangle$ becomes
\begin{align} \label{GaussianMAI}
\chi^{-2}_{\text{MAI}} &= \frac{|\langle [\hat{G}_{\text{opt}},\hat{M}_{\text{MAI}}] \rangle|^2}{\var[\hat{M}_{\text{MAI}}]} \nonumber \\
&= \frac{|\langle 0| \hat{S}^\dagger [\hat{G}_{\text{opt}},\hat{S}\hat{M}_{\text{opt}}\hat{S}^\dagger] \hat{S}|0 \rangle|^2}{\langle 0 | \hat{S}^\dagger \left( \hat{S}\hat{M}_{\text{opt}}\hat{S}^\dagger \right) \left( \hat{S}\hat{M}_{\text{opt}}\hat{S}^\dagger \right) \hat{S} |0 \rangle - \langle 0 | \hat{S}^\dagger \left( \hat{S}\hat{M}_{\text{opt}}\hat{S}^\dagger \right)  \hat{S} |0 \rangle^2 } \nonumber \\
&= \frac{ | \langle 0| \hat{S}^\dagger \hat{G}_{\text{opt}} \hat{S} \hat{M}_{\text{opt}}|0 \rangle-\langle 0 |\hat{M}_{\text{opt}}\hat{S}^\dagger \hat{G}_{\text{opt}} \hat{S} |0 \rangle  |^2  }{\langle 0 | \hat{M}_{\text{opt}}^2 |0 \rangle - \langle 0 | \hat{M}_{\text{opt}} |0 \rangle^2 } \nonumber \\
&= \frac{ |e^r\langle 0 | [\hat{G}_{\text{opt}},\hat{M}_{\text{opt}}]|0 \rangle |^2  }{1/2} \nonumber \\
&= 2e^{2r} \;.
\end{align}
Perhaps surprisingly, we have obtained the same result as Eq.~\eqref{eq:sensGLinM}, meaning $\chi^{-2}=\chi^{-2}_{\text{MAI}}$. This shows that for pure Gaussian states, the MAI protocol does not actually provide any increase in the sensitivity. 
The same conclusion $\chi^{-2}=\chi^{-2}_{\text{MAI}}$ can be derived for mixed Gaussian states Eq.~\eqref{eq:rhoG}, since a simple calculation gives
\begin{align}
\chi^{-2} = \frac{2e^{2r}}{(1+2 \overline{n}_T)} \;,
\end{align}
\begin{align}
\chi_{\text{MAI}}^{-2} = \frac{2e^{2r}}{(1+2\overline{n}_T)} \;,
\end{align}
where we used the thermal state variance $\var[\rho_T,\hat{M}]=(1+2\overline{n}_T)/2$.

\section{MAI advantage in the presence of detection noise}

In the presence of noise in the detection, the measurement of $M$ can be written as $\tilde{M} = \op{M} +\Delta M$, where $\Delta M$ is a random variable describing the noise. This variable follows the Gaussian distribution with mean $\langle \Delta M \rangle =0$ and variance $\langle (\Delta M)^2 \rangle =\sigma^2$. 

In the typical scenario without MAI, the sensitivity to displacements for linear quadrature measurements with detection noise is
\begin{align}
\chi^{-2} &= \frac{|\langle [\hat{G}_{\text{opt}},\tilde{M}_{\text{opt}}] \rangle|^2}{\var[\rho,\tilde{M}_{\text{opt}}]} \nonumber \\
&= \frac{ |\langle [\hat{G}_{\text{opt}},\hat{M}_{\text{opt}}] \rangle  + \langle [\hat{G}_{\text{opt}},\Delta M] \rangle|^2  }{ \langle \left( \hat{M}_{\text{opt}}+\Delta M \right)^2  \rangle - \langle \hat{M}_{\text{opt}}+\Delta M   \rangle^2 } \nonumber \\
&= \frac{ |\langle [\hat{G}_{\text{opt}},\hat{M}_{\text{opt}}] \rangle|^2  }{ \left( \langle\hat{M}_{\text{opt}}^2 \rangle + 2\langle\hat{M}_{\text{opt}} \rangle \langle \Delta M \rangle +\langle \Delta M^2 \rangle \right) - \left( \langle \hat{M}_{\text{opt}} \rangle +\langle\Delta M   \rangle \right)^2 } \nonumber \\
&= \frac{1}{\var[\rho,\op{M}_{\text{opt}}]+\sigma^2 } \;. \label{eq:noisyLIN}
\end{align}

In the scenario with MAI protocol, the measurement with detection noise can be expressed as $\tilde{M}_{\text{MAI}}= \hat{U} (\op{M}+\Delta M) \hat{U}^\dagger = \op{M}_{\text{MAI}}+\Delta M  $  with $\hat{U}=e^{-i\hat{H}t}$, giving the sensitivity
\begin{align}
\chi_{\text{MAI}}^{-2} &= \frac{|\langle [\hat{G}_{\text{opt}},\tilde{M}_{\text{MAI}}] \rangle|^2}{\var[\rho,\tilde{M}_{\text{MAI}} ]} \nonumber \\
&= \frac{|\langle [\hat{G}_{\text{opt}},\hat{M}_{\text{MAI}}] \rangle  + \langle [\hat{G}_{\text{opt}},\Delta M] \rangle|^2 }{ \langle 0| \hat{U}^\dagger \hat{U} \tilde{M}_{\text{opt}} \hat{U}^\dagger \hat{U} \tilde{M}_{\text{opt}} \hat{U}^\dagger  \hat{U} |0\rangle - \langle 0| \hat{U}^\dagger \hat{U} \tilde{M}_{\text{opt}} \hat{U}^\dagger \hat{U} |0\rangle^2 } \nonumber \\
&= \frac{|\langle [\hat{G}_{\text{opt}}, \hat{M}_{\text{MAI}}] \rangle|^2 }{ \langle 0|  \tilde{M}_{\text{opt}}^2  |0\rangle - \langle 0| \tilde{M}_{\text{opt}} |0\rangle^2 } \nonumber \\
&= \frac{|\langle [\hat{G}_{\text{opt}}, \hat{M}_{\text{MAI}}] \rangle_{\rho}|^2 }{ \var[\rho_0,\op{M}_{\text{opt}}] +\sigma^2 } \;. \label{eq:noisyMAI}
\end{align}

Following the results presented in Section IV, Eqs.~(\ref{eq:noisyLIN},\ref{eq:noisyMAI}) can be computed analytically in the Gaussian case, \ie when $K=\Delta=0$ in $\op{H}$, such that $r = 2\epsilon t$. For a pure Gaussian state, since the minimum variance is $\var[\rho,\hat{M}]=e^{-2 r } /2$, the linear sensitivity is 
\begin{equation}
\chi^{-2} = \frac{1}{\frac{1}{2} e^{-2r}+\sigma^2} \;.
\end{equation}
To calculate the MAI sensitivity, using Eq.~\eqref{GaussianMAI} we have 
\begin{equation}
\chi_{\text{MAI}}^{-2} = \frac{e^{2r}}{\frac{1}{2}+\sigma^2 } \;.
\end{equation}

We thus have that, in the presence of detection noise, the improvement in sensitivity due to the MAI protocol is
\begin{equation}
    \dfrac{\chi_{\text{MAI}}^{-2}}{\chi^{-2}}= \dfrac{1+2e^{2r}\sigma^2}{1+2\sigma^2} \;.
\end{equation}
This ratio is grater than one for $\sigma^2>0$ and $r >0$, and in the case of $\sigma^2 \gg 1$ is $e^{2r}$. Again, as pointed out in Section IV, in the absence of detection noise $\sigma=0$ and there is no advantage in using a MAI protocol.

For mixed Gaussian states, the sensitivities in the presence of detection noises are
\begin{align}
\chi^{-2} &= \frac{1}{\frac{(1+2\overline{n}_T)}{2} e^{-2r} +\sigma^2 } \;, \\
\chi_{\text{MAI}}^{-2} &= \frac{e^{2r}}{ \frac{(1+2\overline{n}_T)}{2} +\sigma^2 } \;.
\end{align}
The same conclusion as for the pure state case can be can be obtained by looking at the ratio
\begin{align}
    \dfrac{\chi_{\text{MAI}}^{-2}}{\chi^{-2}}= \dfrac{1+2\overline{n}_T+2e^{2r}\sigma^2}{1+2\overline{n}_T+2\sigma^2} \geq 1\;.
\end{align}
%

% \clearpage
% \newpage

\section{QFI scaling}

\begin{figure}[t]
    \begin{center}
	\includegraphics[width=\textwidth]{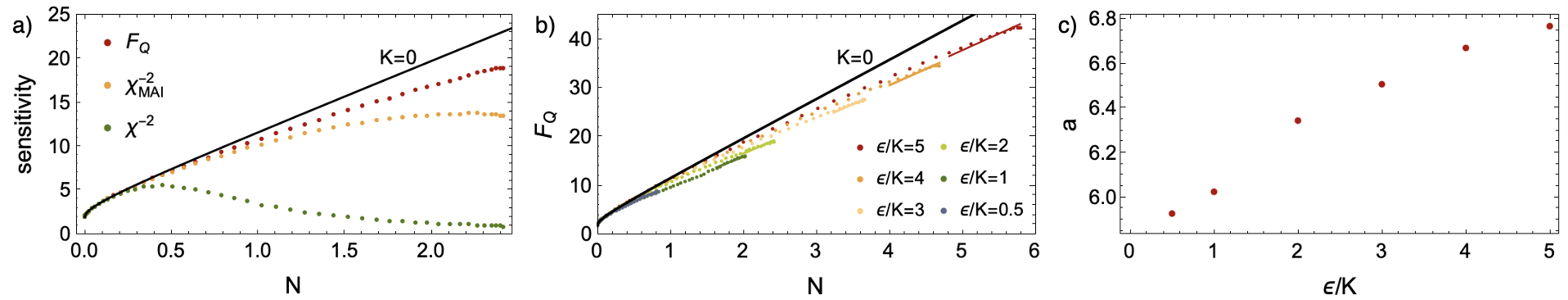}
	\end{center}
    \caption{\textbf{Sensitivity scaling with \bm{$N$}.} Panel a): Sensitivities $F_Q$, $\chi_{\text{MAI}}^{-2}$, $\chi^{-2}$ as functions of the average number of excitations $N$ for the same parameters as in Fig.~4 in the main text, namely $\epsilon/K=2$ and $\Delta/K=0$. 
    Panel b): $F_Q$ as a function of $N$ for different $\epsilon/K$ and $\Delta/K=0$. Here, data points are truncated when the first maximum of $F_Q$ is reached. Solid lines are a linear fit of the last 20\% of points with the model $F_Q=aN+4$.
    Panel c): dependence of the scaling coefficient $a$ on $\epsilon/K$. In the limit $\epsilon/K\rightarrow\infty$ we expect $a=8$.}
    \label{SI_QFIVsN}
\end{figure}
Here, we are interested in investigating how $F_Q$, $\chi_{\text{MAI}}^{-2}$ and $\chi^{-2}$ scale with the average number of excitations $N=\avg{\op{a}^\dagger \op{a}}$ in the system.
Contrary to a simple squeezing transformation, where the one-to-one correspondence between interaction time and $N$ allows us to derive the scaling Eq.~\eqref{eq:scalingSqQFI}, the presence of a Kerr nonlinearity in our system significantly complicates the scaling analysis. When the vacuum state $\ket{0}$ evolves according to the squeezed-Kerr Hamiltonian Eq.~(1) in the main text, given values of $\Delta/K$ and $\epsilon/K$ result in a maximum average number of excitations. Moreover, the evolution of $N$ is in general not monotonic in time. 

Taking as an example the scenario presented in Fig.~4 in the main text, we plot in Fig.~\ref{SI_QFIVsN}a the sensitivities as a function of $N$. For this, note that we stop plotting points after the maximum $N$ (\ie $N\approx 2.4$) is reached. We can see that the QFI scales similarly to the ideal squeezing case (black line, \ie Eq.~\eqref{eq:scalingSqQFI}), and that the scaling of MAI is also comparable. On the other hand, the sensitivity obtained from a linear estimate does not show a favorable scaling with $N$.

The fact that for the scenario we are considering $N$ cannot be arbitrary large significantly complicates the study of these scalings.
Nevertheless, in order to gain some insight, we proceed in the following way. 
We first set a value for $\epsilon/K$, and fix $\Delta/K=0$ for simplicity. Then, we compute QFI and $N$ for the states $e^{-iHt}\ket{0}$ at many different times $t$. We repeat this step for several different $\epsilon/K$, and plot the results in Fig.~\ref{SI_QFIVsN}b. The last 20\% of points before the maximum QFI value are fitted with the model $aN+4$, with free parameter $a$. This is motivated by the fact that the QFI scaling for an ideal squeezed state is $8N+4$, Eq.~\eqref{eq:scalingSqQFI}. The values of $a$ extracted for different values of $\epsilon/K$ are plotted in Fig.~\ref{SI_QFIVsN}c, which shows how the scaling depends on the squeezing rate and Kerr nonlinearity. We expect the trend to saturate at $a=8$ in the limit $\epsilon/K\rightarrow\infty$.

\section{Robustness to losses}

\begin{figure}[t]
    \begin{center}
	\includegraphics[width=8cm]{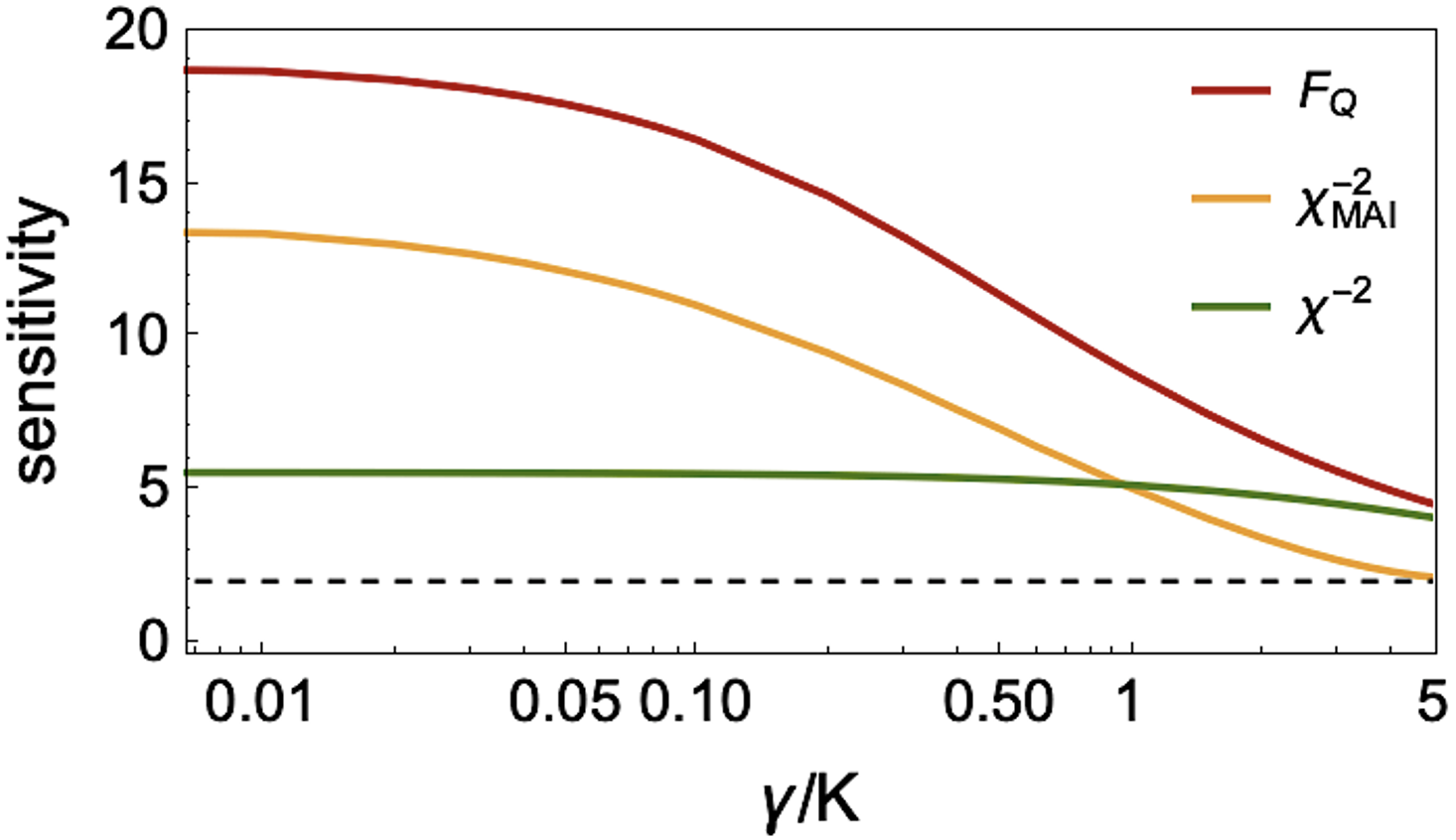}
	\end{center}
    \caption{\textbf{Robustness to losses.} Maximum of the sensitivities $F_Q$, $\chi_{\text{MAI}}^{-2}$ and $\chi^{-2}$ as a function of energy relaxation rate $\gamma/K$, for the same parameters as in Fig.~4 in the main text. The dashed black line represents the standard quantum limit.}
    \label{SI_lossrate}
\end{figure}

In the analysis of our metrological protocol, we were interested in making sure that MAI method was giving an advantage also in the presence of losses, which are inevitable experimentally. Our simulations showed that high levels of losses are required before the MAI sensitivity would fall below the shot noise limit, or below the linear measurements sensitivity, see Fig.~4 of the main text. In this sense, we say that the MAI method is robust to losses.

To further investigate how $F_Q$, $\chi_{\text{MAI}}^{-2}$ and $\chi^{-2}$ depend on the level of losses, we look at how the maxima of these quantities change as a function of $\gamma/K$. This is plotted in Fig.~\ref{SI_lossrate}. Even if the linear measurements sensitivity, $\chi^{-2}$, is almost unaffected by the amount of losses considered, note that this is still surpassed by $\chi_{\text{MAI}}^{-2}$ for a wide range of experimentally relevant $\gamma/K$. We thus conclude that the MAI method can indeed provide an advantage over the linear measurement in concrete experimental scenarios.

\end{widetext}

\end{document}